\documentstyle[11pt,amssymb,epsf,cite]{article}

\makeatletter
\@addtoreset{equation}{section}
\makeatother

\def\ben{\begin{equation}}
\def\een{\end{equation}}

\let\a=\alpha    
    
 \let\m=\mu \let\n=\nu

\let\C=\Chi

\def\nn{\nonumber} \def\bd{\begin{document}} \def\ed{\end{document}}
\def\ds{\documentstyle} \let\fr=\frac \let\bl=\bigl \let\br=\bigr
\let\Br=\Bigr \let\Bl=\Bigl
\let\bm=\bibitem
\let\na=\nabla
\let\pa=\partial \let\ov=\overline
\newcommand{\be}{\begin{equation}}
\newcommand{\ee}{\end{equation}}
\def\ba{\begin{array}}
\def\ea{\end{array}}
\def\ft#1#2{{\textstyle{{\scriptstyle #1}\over {\scriptstyle #2}}}}
\def\fft#1#2{{#1 \over #2}}
\def\del{\partial}
\def\vp{\varphi}
\def\sst#1{{\scriptscriptstyle #1}}
\def\oneone{\rlap 1\mkern4mu{\rm l}}
\def\td{\tilde}
\def\wtd{\widetilde}
\def\ie{\rm i.e.\ }
\def\dalemb#1#2{{\vbox{\hrule height .#2pt
        \hbox{\vrule width.#2pt height#1pt \kern#1pt
                \vrule width.#2pt}
        \hrule height.#2pt}}}
\def\square{\mathord{\dalemb{6.8}{7}\hbox{\hskip1pt}}}
\newcommand{\ho}[1]{$\, ^{#1}$}
\newcommand{\hoch}[1]{$\, ^{#1}$}
\newcommand{\bea}{\begin{eqnarray}}
\newcommand{\eea}{\end{eqnarray}}
\newcommand{\ra}{\rightarrow}
\newcommand{\lra}{\longrightarrow}
\newcommand{\Lra}{\Leftrightarrow}
\newcommand{\ap}{\alpha^\prime}
\newcommand{\bp}{\tilde \beta^\prime}
\newcommand{\tr}{{\rm tr} }
\newcommand{\Tr}{{\rm Tr} }
\def\0{{\sst{(0)}}}
\def\1{{\sst{(1)}}}
\def\2{{\sst{(2)}}}
\def\3{{\sst{(3)}}}
\def\4{{\sst{(4)}}}
\def\5{{\sst{(5)}}}
\def\6{{\sst{(6)}}}
\def\7{{\sst{(7)}}}
\def\8{{\sst{(8)}}}
\def\n{{\sst{(n)}}}
\def\cA{{{\cal A}}}
\def\cF{{{\cal F}}}
\def\tV{\widetilde V}
\def\tW{\widetilde W}
\def\tH{\widetilde H}
\def\tE{\widetilde E}
\def\tF{\widetilde F}
\def\tA{\widetilde A}
\def\im{{{\rm i}}}
\def\tY{{{\wtd Y}}}
\def\ep{{\epsilon}}
\def\vep{{\varepsilon}}
\def\R{\rlap{\rm I}\mkern3mu{\rm R}}
\def\bD{{{\bar D}}}

\def\R{\rlap{\rm I}\mkern3mu{\rm R}}
\def\bD{{{\bar D}}}
\def\R{{{\Bbb R}}}
\def\C{{{\Bbb C}}}
\def\H{{{\Bbb H}}}
\def\CP{{{\Bbb C}{\Bbb P}}}
\def\RP{{{\Bbb R}{\Bbb P}}}
\def\Z{{{\Bbb Z}}}
\def\bA{{{\Bbb A}}}
\def\bB{{{\Bbb B}}}
\def\bC{{{\Bbb C}}}
\def\bR{{{\Bbb R}}}
\def\bD{{{\Bbb D}}}
\def\bE{{{\Bbb E}}}
\def\bZ{{{\Bbb Z}}}
\def\cD{{{\cal D}}}
\def\Re{{{\frak{Re}}}}
\def\Im{{{\frak{Im}}}}
\def\cosec{{\,\hbox{cosec}\,}}
\def\Gm{{\Gamma_{\!\! -}}}
\def\Gp{{\Gamma_{\!\! +}}}

\def\stan{{standard }}
\def\nonstan{{supernumerary }}

\def\cosech{{\hbox{cosech}}}

\def\etcyc{{\hbox{and cyclic}}}
\def\btheta{{\bar\theta}}

\newcommand{\tamphys}{\it Center for Theoretical Physics,
Texas A\&M University, College Station, TX 77843, USA}
\newcommand{\umich}{\it Michigan Center for Theoretical Physics,
University of Michigan\\ Ann Arbor, MI 48109, USA}
\newcommand{\upenn}{\it Department of Physics and Astronomy,\\
University of Pennsylvania, Philadelphia,  PA 19104, USA}
\newcommand{\SISSA}{\it  SISSA-ISAS and INFN, Sezione di Trieste\\
Via Beirut 2-4, I-34013, Trieste, Italy}

\newcommand{\mitchell}{\it George P. \& Cynthia W.
Mitchell Institute for Fundamental Physics,\\
Texas A\&M University, College Station, TX 77843-4242, USA}

\newcommand{\newton}{\it Isaac Newton Institute for Mathematical Sciences,\\
0 Clarkson Road,  University of Cambridge,
Cambridge CB3 0EH, UK}

\newcommand{\ihp}{\it Institut Henri Poincar\'e\\
  11 rue Pierre et Marie Curie, F 75231 Paris Cedex 05}

\newcommand{\damtp}{\it DAMTP, Centre for Mathematical Sciences,
 Cambridge University,\\  Wilberforce Road, Cambridge CB3 OWA, UK}

\newcommand{\itp}{\it Institute for Theoretical Physics, University of
California\\ Santa Barbara, CA 93106, USA}

\newcommand{\istanbul}{\it Department of Mathematics,  Bo{\u g}azi{\c c}i
University, Bebek, Istanbul 34342, Turkey.}

\newcommand{\gursey}{\it Feza Gursey Institute, Cengelkoy, 81220,
Istanbul, Turkey}

\newcommand{\CTP}{\it Center for Theoretical Physics, Laboratory for
Nuclear Science and Department of Physics, Massachusetts Institute
of Technology, Cambridge, MA 12139 USA}

\newcommand{\auth}{
M. Cariglia\hoch{\star}, G.W. Gibbons\hoch{\star\flat}, R. G\"uven\hoch{\sharp} and
C.N. Pope\hoch{\ddagger}}

\begin{document}
\begin{flushright}
\hfill {
DAMTP-2003-140\ \ \
MIFP-03-24\ \ \
MIT-CTP-3451\ \ \
}\\
\hfill{
\bf hep-th/0312256}
\end{flushright}

\begin{center}

{\Large {\bf Non-Abelian pp-waves in $D=4$ supergravity theories }}

\vspace{12pt}

\auth

\vspace{7pt}

{\hoch{\star}\damtp}

\vspace{7pt}
{\hoch{\sharp}\istanbul}

\vspace{7pt}
{\hoch{\ddagger}\mitchell}

\vspace{7pt}
{\hoch{\flat}\CTP}

\vspace{14pt}

\underline{ABSTRACT}
\end{center}
The non-Abelian plane waves, first found in flat spacetime by
Coleman and subsequently generalized to give pp-waves in
Einstein-Yang-Mills theory, are shown to be ${1 \over 2}$-
supersymmetric solutions of a wide variety of $N=1$ supergravity
theories coupled to scalar and vector multiplets, including the
theory of $SU(2)$ Yang-Mills coupled to an axion $\sigma$ and
dilaton $\phi$ recently obtained as the reduction to
four-dimensions of the six-dimensional Salam-Sezgin model. In this
latter case they provide the most general supersymmetric solution.
Passing to the Riemannian formulation of this theory we show that
the most general supersymmetric solution may be constructed
starting from a self-dual Yang-Mills connection on a self-dual
metric and solving a Poisson equation for $e^\phi$. We also
present the generalization of these solutions to non-Abelian AdS
pp-waves which allow a negative cosmological constant and preserve
${1 \over 4}$ of supersymmetry.

{\vfill\leftline{}\vfill \vskip 10pt \footnoterule {\footnotesize
\hoch{\ddagger} Research supported in part by DOE grant
DE-FG03-95ER40917

\vskip 10pt {\footnotesize\hoch{\sharp} Research supported in part
by the Turkish Academy of Sciences (TUBA) \vskip -12pt} \vskip
14pt }

\pagebreak
\setcounter{page}{1}

\newpage

\section{Introduction}

There is a considerable literature on Yang-Mills theory coupled
to Einstein gravity in four-dimensions,
 but comparatively few exact results are known outside the realm
of Abelian embeddings, where the problem reduces to the even more
intensively studied  Einstein-Maxwell theory. In the
four-dimensional flat Minkowski spacetime, Coleman \cite{coleman}
exhibited genuinely non-abelian solutions of plane wave type and
these were subsequently generalized to incorporate the effects of
gravity \cite{guv},\cite{traut}. Non-abelian plane waves were then
embedded into ten-dimensional Superstring Theory in \cite{Guv2}
but the Yang-Mills field considered there did not arise through a
process of dimensional reduction. It was found that these
solutions constitute a natural generalization of the familiar
vacuum pp-wave spacetimes. In the particular case of the
non-Abelian plane waves it was also noted that, as in the case of
the four-dimensional vacuum plane-waves \cite{gwg} \cite{des}, the
solutions are not affected by quantum corrections. In the
Riemannian regime a great deal is known about self-dual Yang-Mills
solutions on a fixed background, but these are not usually
consistent unless the metric is Ricci flat. One particularly
interesting case is the extension of the well-known 't Hooft
ansatz to a Hyper-K\"ahler background apparently first written
down in \cite{aragone} (see \cite{EtesiHausel} \cite{Etesi} for
recent discussions).
 There is also a considerable literature on Einstein-Yang-Mills
theory coupled to a dilaton, but again few exact results are known.

For physical interpretations it is natural to view these problems
from the point of view of supersymmetry and of higher dimensions.
The smallest dimension which yields a non-Abelian Yang-Mills
theory through the Kaluza-Klein reduction is six and recently, two
of us have shown \cite{gibpop} how the $SU(2)$ Yang-Mills fields
together with an axion and a dilaton, coupled to the
four-dimensional Einstein gravity can be obtained as a consistent
and supersymmetric Pauli type reduction of the six-dimensional
$(1,0)$ gauged supergravity theory. This theory was first
constructed by Nishino and Sezgin \cite{nissez} and its reduction
to four dimensions was first discussed by Salam and Sezgin
\cite{salsez}.  The resulting four-dimensional theory
 contains a scalar multiplet whose bosonic sector comprises
an axion and a dilaton together with an $SU(2)$  Yang-Mills vector
multiplet. There is no potential for the scalars. The theory can
be seen as a special case of the general coupling of scalar and
vector multiplets to $N=1$ supergravity first  worked out by
Cremmer et
al.\cite{Cremmer_et_al_1,Cremmer_et_al_2,Cremmer_et_al_3}. Under
some circumstances, for example for purely magnetic or purely
electric solutions, the axion field $\sigma$ can  be consistently
set to a constant value in which case the equations of motion
coincide with those of the much studied  Einstein-Maxwell-dilaton
system modulo the non-Abelian nature of the spin-1 fields. An
interesting class of non-trivial Abelian solutions are of
Robinson-Trautman type \cite{Robinson_Trautman}.

    In the study of all supersymmetric backgrounds of the six-dimensional
gauged supergravity theory a sub-problem is to find all
supersymmetric solutions of its four-dimensional Pauli reduction.
One of the purposes of the present letter is to give the complete
solution of this sub-problem and this will lead to an interesting
connection to the old works on non-Abelian pp-waves. We shall show
that the non-Abelian pp-waves exhaust the class of all
supersymmetric solutions of the four-dimensional equations of the
Salam-Sezgin model. From this result it follows that Coleman's
original solutions exhaust the class of supersymmetric Yang-Mills
fields in flat spacetime thus giving them a possible significance.
Moreover, we shall show that non-Abelian pp-waves are in fact
supersymmetric solutions of the much wider class of theories
constructed by Cremmer et al. We shall also furnish the
generalization of the non-Abelian pp-wave solutions which
incorporates a negative cosmological constant.  These will be the
non-Abelian anti-de Sitter (AdS) pp-waves which may have potential
applications to the AdS/CFT correspondence. We shall begin by
considering the simplest Salam-Sezgin model.

\section{The Salam-Sezgin Model}

   In a previous letter, three of us showed \cite{ggp} that the
 remarkable supersymmetric background found by Salam and Sezgin is in
 fact unique among all non-singular solutions of the six-dimensional
 theory with four-dimensional Poincar\'e, de Sitter or anti-de Sitter
 invariance. The present section is to some extent complementary: we
 obtain all supersymmetric solutions of the four-dimensional reduced
 theory. The reduced four-dimensional theory describes $N=1$
 supergravity coupled to a scalar multiplet whose bosonic sector
 comprises an axion $\sigma$ and dilaton $\phi$ and an $SU(2)$ vector
 multiplet whose bosonic sector consists of a Yang-Mills connection $
 A^i$.

    The bosonic Lagrangian of this reduced theory is
\be
{\cal L} = R\, {*\oneone} - \ft12 {*d\phi}\wedge d\phi
-\ft12 e^{2\phi}\, {*d\sigma}\wedge d\sigma - \ft12
e^{-\phi}\, {*F^i}\wedge F^i + \ft12 \sigma\, F^i\wedge F^i\,,
\label{4lag2}
\ee
and the
four-dimensional fermionic supersymmetry transformations are
\bea
\delta\lambda^i &=& -\ft1{4\sqrt2}\, e^{-\ft12\phi}\,
F^i_{\a\beta}\, \gamma^{\a\beta}\ep\,, \label{vector}
\eea
\bea
\delta\chi &=& \ft14 (\del_\a\phi\, -\im\, e^\phi\, \del_\a \sigma\,
\gamma_5) \gamma^\a\, \ep \,,\label{spinor}
\eea
\bea
\delta\psi_\a' &=& \nabla_\a\, \ep  -\ft{\im}{4}\,
e^\phi\, \del_\a\sigma\, \gamma_5\ep\,.\label{gravitino}
\eea

   Since we are using spacetime signature $(-,+,+,+)$, there exists a
representation of the gamma matrices such that they are all real.  Our
convention for $\gamma_5$ is such that $\gamma_5^2=+1$, and hence
$\im\, \gamma_5$ is also real in this representation.  Thus we may
regard the spinors of the four-dimensional theory either as purely
real (\ie Majorana) or as complex and Weyl.

   Note that if $F^i\wedge F^i=0$, then it is consistent to set the
field $\sigma$ to a constant value, and the equations of motion
reduce to those of the pure Einstein-Yang-Mills-Dilaton system.

   The consistent embedding of the above four-dimensional $N=1$ theory into
the six-dimensional chiral supergravity of Salam and Sezgin was obtained in
\cite{gibpop}; in the bosonic sector, the reduction ansatz is given by
\bea
d\hat s_6^2 &=& e^{\ft12\phi}\, ds_4^2 + e^{-\ft12 \phi}\, g_{mn}\,
(dy^m + 2g\, K^m_i\, A^i)(dy^n + 2g\, K^n_j\, A^j)\,,\nn\\
\hat F_\2 &=& \fft1{2g}\, \Omega_2
   -d(\mu^i\, A^i)\,,\nn\\
\hat H_\3 &=& H_\3 - 2g\,  F^i\wedge K_{i\, m}\,(dy^m + 2g\, A^j\, K^m_j)
\,,\nn\\
\hat \phi &=& -\phi\,,
 \eea
where $K^m_i= (8g^2)^{-1}\, \ep^{mn}\, \del_n\, \m^i$ are the Killing
vectors on the unit 2-sphere defined by $\mu^i\, \mu^i=1$, and
$g_{mn}$ is the metric on the 2-sphere scaled to radius $1/(2\sqrt 2\,
g)$. If this ansatz is substituted into the equations of motion
following from the Lagrangian
\be
{\cal L}= \hat R\, {\hat *\oneone} - \ft14 {\hat *d\hat \phi}\wedge d\hat\phi
- \ft12 e^{\hat\phi}\, {\hat* \hat H_\3}\wedge \hat H_\3 - \ft12
e^{\ft12\hat\phi}\, {\hat* \hat F_\2}\wedge \hat F_\2 -
8g^2\, e^{-\ft12\hat\phi}\, {\hat *\oneone}
\ee
of the Salam-Sezgin theory, where $\hat F_2 =d\hat A_\1$, $\hat H_\3 =
d\hat B_\2 + \ft12 \hat F_\2\wedge \hat A_\1$, one obtains the
equations of motion following from (\ref{4lag2}), where $H_\3 =
e^{2\phi}\, {*d\sigma}$ \cite{gibpop}.

\subsection{The Lorentzian Theory \label{sec:Lorentzian}}

    In order to maintain the supersymmetry in the bosonic sector,
the spacetime must admit Killing spinors for which the variations
(\ref{vector})-(\ref{gravitino}) of the fermionic fields are zero.
Assuming that $\epsilon$ is a commuting Majorana spinor, the only
nontrivial bilinears  of $\epsilon$ are $ K_{\mu} =
{\bar{\epsilon}} \gamma_{\mu} \epsilon$ and $ {\bar{\epsilon}}
\gamma_{\mu \nu} \epsilon$ and it turns out that the vector
$K_{\mu}$ can be used to completely characterize the solutions.
From the vanishing of the right hand side of (\ref{gravitino}) it
can be deduced that $K_{\mu}$ must be a covariantly constant null
vector:
 \be
  \nabla_{\mu} K_{\nu} =0, \hspace{1.5 cm} K^{\mu}
K_{\mu} = 0,
 \ee
and consequently, the metric must be that of a
pp-wave (see \cite{MacCallum}, pg.380-381):
 \bea ds^2 = 2dudv + H(u,x,y) du^2 + dx^2 + dy ^2.
\label{eq:pp_metric}
 \eea
 The vanishing of the right hand side of (\ref{vector}) implies that
\bea
 F^i_{\mu \nu} K^\nu = \star F^i_{\mu \nu} K^\nu = 0,
\label{elecmag}
 \eea
where $\star$ is the Hodge dual. Defining the basis one-forms as
$e^+= du$, $e^- = dv + \ft12 H\, du$, $e^a = dx^a$, where
$x^a=(x,y)$ and $ds^2 = 2 e^+\, e^- + e^a\, e^a$, the
non-vanishing frame components of the torsion-free spin connection
and the curvature are given by
\be
\omega_{+a} = \ft12 \del_a H\, e^+ \,,\qquad R_{+a+b} = -\ft12
\del_a \del_b H\,,\qquad
R_{++} = -\ft12 \nabla^2\, H\,.\label{omr}
\ee
The generalization of Coleman's non-Abelian plane waves
\cite{coleman} to curved spacetime  now shows that there is a
gauge in which the Yang-Mills potential can be chosen as
\cite{guv}
\bea A^i= {\cal A} ^i (u,x,y)  \,du\,.
 \label{eq:A_gauge} \eea
In this gauge the Yang-Mills field strength is $F^i = \del_a{\cal
A}^i\, dx^a\wedge du$. Notice that although the nonlinear term $A
\wedge A$ in the Yang-Mills curvature vanishes the solutions will
have a full non-Abelian character as long as  the holonomy of the
Yang-Mills connection is not contained in any $U(1)$ subgroup of
$SU(2)$. The vanishing of $A \wedge A$ just reflects the fact that
the non-Abelian pp-waves constitute their own linearized
approximation. This is a well-known property of the vacuum
pp-waves.

    Using the pp-wave curvature given in (\ref{omr})
it can be inferred from (\ref{gravitino}) that $\gamma^+\, \ep=0$
and  $ d(e^{\phi} d\sigma) \gamma_5 \epsilon = 0$.  The second
condition allows us to write $ e^\phi d\sigma = d\lambda$, where
$\lambda$ is an arbitrary function. On the other hand, since
(\ref{spinor}) implies $K^{\mu} \del_{\mu}\phi =K^{\mu}
\del_{\mu}\sigma = 0$, one can deduce from (\ref{spinor}) that the
dilaton $\phi$ and the axion $\sigma$ can only be arbitrary
functions of $u$. Therefore, $\lambda = \lambda(u)$.

   It remains to check whether the field equations are satisfied.
Since $\phi = \phi(u)$,  $\sigma = \sigma(u)$ and the scalar
invariants of the Yang-Mills curvature vanish, the dilaton and the
axion field equations are trivially satisfied. The Yang-Mills
equations reduce to
\bea
(\partial ^2_x+ \partial ^2_y) {\cal A}  ^i =0\,.
\label{eq:YM_eqs}
\eea
Defining $z=x+iy$, we may solve this equation by writing
\bea {\cal A}^i={1 \over 2} \big [ \chi^i(u,z) +{\bar \chi}^i(u,
{\bar z}) \big ]\,, \label{achi} \eea
where the $su(2)$ valued functions $\chi^i(u,z)$ are holomorphic
in $z$, and arbitrary in $u$.

 The only non-vanishing frame
component of the energy-momentum tensor for this field
configuration is $T_{++} =  \ft12[ ({\dot{\phi}}^2 +
{\dot{\lambda}}^2) + e^{-\phi} \del_a {\cal A}^i\, \del_a {\cal
A}^i ]$, where dot denotes the differentiation with respect to
$u$.  Using (\ref{omr}) the Einstein equations become $ -4
\del\bar\del H = ({\dot{\phi}}^2 + {\dot{\lambda}}^2 ) + e^{-\phi}
\del\chi^i\, \bar\del\bar\chi^i$, where $\del=\del/\del z$ and
$\bar\del = \del/\del\bar z$.  It follows that the metric function
$H(u,z,{\bar z})$ is given by
\bea H(u,z,{\bar z}) =K(u, z) + {\bar K} (u, {\bar z})- \ft14 [
e^{-\phi}{ \chi} ^i(u,z) {\bar\chi} ^i(u, {\bar z})
+({\dot{\phi}}^2 + {\dot{\lambda}}^2) z \bar{z} ]\,, \eea
where $K(u,z)$ is an arbitrary function of $u$ but holomorphic in $z$.

    With $\phi$ and $\sigma$ arbitrary functions of $u$, it can
be checked  that the supersymmetry constraints from the
transformation rules (\ref{vector}) - (\ref{gravitino}) are now
all satisfied except the condition
\be
    d\epsilon = \ft{\im}4\, d\lambda \, \gamma_5 \epsilon,
\ee
 which follows from (\ref{gravitino}). This equation can be
integrated to fix the final form of the Killing spinor:
 \be
 \epsilon = e^{\ft{\im}4\, \lambda(u) \, \gamma_5} \, \eta\,,
 \ee
where  $d\eta = 0$. The constant spinor $\eta$ obeys
 \be
 \gamma^+\eta = 0,
 \ee
 showing that the solutions possess 1/2 of the Poincar\'e
 supersymmetry.

\subsection{The Riemannian  Theory}

    This is most easily obtained by setting $\sigma =i\Sigma$, with
$\Sigma$ real and inserting the appropriate factor of $i$ in the
$F^i\wedge F^i$ term in the Lagrangian. The scalar fields
$\phi,\Sigma$ then take their values in two-dimensional de-Sitter or
Anti-de-Sitter space rather than two-dimensional hyperbolic space or
the upper-half- plane as it does for the Lorentzian theory. The
indefinite metric on the target space is
\bea
d \phi^2 -e^{2 \phi} d \Sigma ^2.
\eea

    In the Euclidean-signatured theory there are no Majorana spinors,
and so we take the spinors to be chiral.   From (\ref{vector}),
and choosing $\ep= +\gamma_5 \ep$, we see that supersymmetry then
requires that $F^i $ must be anti-self-dual
\bea
 F^i=-\star F^i.
\eea
We easily see from (\ref{spinor}) that
\bea
\partial _\mu \phi + e^\phi \partial _\mu \Sigma =0. \label{lightlike}
\eea
The geometrical meaning of (\ref{lightlike}) is that the image
of the map $\phi(x), \Sigma (x)$
in the two-dimensional internal space is a
 lightlike geodesic in de-Sitter or
anti-de-Sitter spacetime.
From (\ref{gravitino}) we deduce that
\bea
\ep = e^{- {1 \over 4} \phi } \eta,
\eea
where $\eta$ is covariantly constant. It follows that the
metric must be anti-self-dual or Hyper-K\"ahler
\bea R_{\mu \nu \alpha \beta }=- {1\over 2} \epsilon _{\mu \nu
\sigma \tau} R^{\sigma \tau }\,_{\alpha \beta} \, .
\eea
In particular the Ricci tensor must vanish. This is consistent
with the field equations because the stress tensor of a self-dual
Yang-Mills field and of a  map with totally light like image
vanish identically. It remains to check the equation of motion for
the scalar fields. A short calculation reveals that it reduces to
the linear equation

\bea
\nabla ^2 e^{\phi} =- { 1\over 4} F^i_{\mu \nu} F^{i \mu \nu},
\label{phieq}
\eea
where $\nabla^2 $ is the scalar Laplace operator
on the Hyper-K\"ahler 4-manifold.

   A simple example is to take the four-dimensional metric to be
flat, which we write as $ds_4^2 = dr^2 + \ft14 r^2\, \sigma_i^2$,
where
 $\sigma_i$ are left-invariant one-forms on $SU(2)$, and
the Yang-Mills fields to be those of the single-instanton solution.  This
can be written as
\be
A^i = \fft1{1+\lambda\, r^2}\, \sigma_i\,,
\ee
where $\lambda$ sets the scale size of the instanton.  It is straightforward
to integrate (\ref{phieq}), yielding
\be
e^\phi = c +\fft{m}{r^2} +
\fft{2\lambda\,  (2+\lambda\, r^2)}{(1+\lambda\, r^2)^2}\,,
\ee
where $c$ and $m$ are arbitrary constants of integration. The solution
is non-singular at $r=0$ if $m=0$.


\section{Non-Abelian AdS pp--waves}\label{AdS}

One can readily generalize the discussion of section
(\ref{sec:Lorentzian}) to the case when a negative cosmological
constant term is present. Consider adding a term $-2 \Lambda
*\oneone $ to the action (\ref{4lag2}), where $ \Lambda$ is
constant. This will bring in a cosmological constant term in
Einstein's equations and the appropriate ansatz for the metric
that modifies (\ref{eq:pp_metric}) is now
\bea ds^2 =  \frac{a^2}{z^2} \left( 2dudv + H(u,x,z) du^2 + dx^2
+dz^2 \right) , \label{eq:AdS_pp_metric} \eea
where the constant $a$ is proportional to the radius of curvature.
Note that in this section $z$ is used to denote a real coordinate,
and should not be confused with the complex coordinate used in
section 2.1. The metric (\ref{eq:AdS_pp_metric}) is related to
(\ref{eq:pp_metric}) simply by a conformal factor. On each
(holographic) slice $z={\rm constant}$, it describes a wave
propagating at the speed of light along the horosphere
\cite{Siklos,Gibbons_Ruback}.

Assuming that $\phi=\phi(u),\,  \sigma=\sigma(u)$ and  using the
gauge (\ref{eq:A_gauge}) for $A^i$, the Yang-Mills equations are
still given by (\ref{eq:YM_eqs}) and admit the same solution as in
section 2.1. Once again the scalar invariants of the Yang- Mills
field vanish and the dilaton and the axion field equations put no
restrictions on the arbitrary functions $ \phi(u)$ and $
\sigma(u)$. The metric (\ref{eq:AdS_pp_metric}) is a solution of
the generalized Einstein equations
\bea R_{\alpha\beta} = \Lambda g_{\alpha\beta} + \frac{1}{2}\,
e^{-\phi} \, \left( F_{\alpha\gamma}^i F^{i\;\;\gamma}_{\beta} -
\frac{1}{4} F_{\gamma\delta}^i F^{i\;\;\gamma\delta} \,
g_{\alpha\beta} \right) + \frac{1}{2}\,{\nabla_{\alpha}}\phi
{\nabla_{\beta}}\phi + \frac{1}{2}\, e^{2\phi}
{\nabla_{\alpha}}\sigma {\nabla_{\beta}}\sigma ,
 \eea
provided that $\Lambda = - 3/ a^2$ and $H$ satisfies a Siklos
equation \cite{Podolsky}  with source:
\bea - ( \partial_x^2 + \partial_z^2 ) H + \frac{2}{z} \partial_z
H = e^{-\phi} \frac{z^2}{a^2} \, (\partial A)^2 + {\dot{\phi}}^2 +
{\dot{\lambda}}^2 ,
 \label{eq:gen_Siklos}
 \eea
where $(\partial A)^2 = \partial_z A^i \partial_z A^i + \partial_x
A^i
\partial_x A^i$ and as in section \ref{sec:Lorentzian}, $d\lambda = e^{\phi}d\sigma$ .
The solutions of (\ref{eq:gen_Siklos}) with zero source are known
to be singular at $z= +\infty$ \cite{Chamblin_Gibbons}.

In view of an AdS/CFT interpretation of the non-Abelian pp--waves
it may be useful to consider the same construction in all
dimensions $D \ge 4$. The metric is
\bea ds^2 =  \frac{a^2}{z^2} \left( 2dudv + H(u,x_m,z) du^2 +
h_{mn} dx^m dx^n + dz^{2}\right) , \eea
where we allow a Ricci flat metric $h_{mn}$ on the
$D-3$-dimensional transverse space. We keep the same functional
dependence of the scalar fields on the coordinates: $\phi=\phi(u)$
and $\sigma=\sigma(u)$. The definition of the Yang-Mills potential
is also unchanged but its equation of motion becomes
\bea ( \nabla_{\perp}^2 + \partial_z^2 - (D-4)
\frac{1}{z}\partial_z ) A^i = 0 ,
\eea
where $\nabla_{\perp}^2$ is the Laplace operator of $h_{mn}$. In
D-dimensions the cosmological constant is  $\Lambda = - (D-1)/
a^2$ and the Siklos equation with source generalizes to
\bea
- \nabla_{\perp}^2 H - z^{D-2} \partial_z \left(
\frac{1}{z^{D-2}} \partial_z H \right)  = e^{-\phi}
\frac{z^2}{a^2} \, (\partial A)^2 + {\dot{\phi}}^2 +
{\dot{\lambda}}^2 .
\eea
where $(\partial{ A})^2 = h^{mn} \partial_{m} A^{j} \partial_{n}
A^{j} + \partial_{z} A^{j} \partial_{z} A^{j} $ and the gauge
group is arbitrary.

\section{Non-Abelian pp-waves in General $D=4$
Supergravities\label{sec:general}}

   In this section we show that the non--Abelian pp--waves
considered above are supersymmetric solutions of the  general
class of the four-dimensional supergravity theories describing
matter with arbitrary potential coupled to Yang-Mills fields. The
model of supergravity interacting with scalar multiplets is given
in \cite{Cremmer_et_al_1}, while the coupling with Yang-Mills
fields was first constructed in
\cite{Cremmer_et_al_2,Cremmer_et_al_3}. The bosonic sector of the
theory is described by the Lagrangian
\begin{eqnarray}
{\cal L} &=& R\, {*\oneone} -  2 \, G^i_{\;\; j} \,  {*D z_i}\wedge
D\overline{z}^j
 -  Re f_{ab}\,  {*F^a}\wedge F^b -  \, Im f_{ab}\,  F^a\wedge F^b
\nonumber \\ &-& g^2 \, Re f^{-1}_{ab} \, G^i \, (T_a)_i^{\;\; j}
\, z_j \, G^k \, (T_b)_k^{\;\; l} \, z_l + 2 \, e^G \, \left( 3 -
G_i \, (G^{-1})^i_{\;\; j} \, G^j \right) {*\oneone} ,
 \label{eq:Cremmer}
\end{eqnarray}
where $z_i= A_i + i \,B_i$ are complex scalar fields in a
realization of the gauge group  whose Yang-Mills field strength
is $F^a_{\mu\nu}$ and $D_\mu z_i$ is the gauge covariant
derivative. Moreover, $g$ denotes the gauge coupling constant,
$G(z,\overline{z})$ is the K\"{a}hler potential, $G^i =
\partial G / \partial z_i$, $G_j =
\partial G / \partial \overline{z}^j$, $G^i_{\;\; j} = \partial^2 G /
\partial z_i \partial \overline{z}^j$ and $f_{ab}(z)$ is a set of
holomorphic functions.

    The Lagrangian (\ref{eq:Cremmer}) ostensibly requires that the
scalar potential is non-zero.  However, we can always add a
constant to the K\"ahler potential $G$ without affecting $G_i$ or
$G^i{}_j$. Taking the constant to minus infinity allows us to
discard the scalar potential term.  In this way, we can view the
dimensionally-reduced Salam-Sezgin Lagrangian (\ref{4lag2}) as a
special case of (\ref{eq:Cremmer}).  Note that in this case the
gauge covariant derivative on the scalar fields becomes the
ordinary derivative, because the $SU(2)$ gauge group does not act
on the scalar fields $(\phi,\sigma)$. The non-Abelian pp-waves of
section \ref{sec:Lorentzian} are therefore solutions in this limit
and it is clear that the same procedure can be employed for other
gauge groups as well.

   In general the fermionic supersymmetry transformations are
\bea
\delta\lambda^a &=& - \frac{1}{2}\, F^a_{\a\beta}\,
\gamma^{\a\beta}\ep_L  + \frac{i}{2}\, g \, Re f^{-1}_{ab} \, G^i
\, (T_b )_i^{\;\; j} \, z_j \ep_L  , \label{eq:gaugino}\\
\delta\chi_i &=& -  \, \gamma^\a D_\alpha  \left( A_i + i \,
\gamma_5 \, B_i \right) \, \ep - \sqrt{2} \, e^{G/2} \,
(G^{-1})_i^{\;\; j} \, G_j \ep  \,,         \label{eq:chi}\\
\delta\psi_{\alpha}  &=& 2\,  \nabla_{\alpha} \, \ep  + \left( G^i \,
D_\alpha z_i - G_i \, D_\alpha \overline{z}^i \right) \ep +  \,
e^{G/2} \, \gamma_\alpha \, \ep \,.\label{eq:gravitino}
\eea
As in section \ref{sec:Lorentzian}, it is easy to see that (\ref{eq:gaugino})
implies
\begin{eqnarray}
    \iota_V F_a &=&  \, Re \, X_a \, V , \\
    \iota_V * \, F_a &=&  \, Im \, X_a \, V ,
\end{eqnarray}
where $X_a =  \frac{i}{2}\, g \, Re f^{-1}_{ab} \, G^i\, (T_b
)_i^{\;\; j} \, z_j$ and $V$ is the null vector associated to the Weyl
spinor $\ep_L$. In order to have a non-abelian pp-wave solution one
has to impose then $X_a = 0$, which implies $G^i =0$. This
cannot be identically true otherwise there would be no kinetic term
for the scalars. Rather, one can ask the scalars to be at an extremal
point of the K\"{a}hler function
\begin{eqnarray}
    G^i|_{z=z_0} &=& 0 , \label{eq:extremal_1}  \\
    D_\alpha z_i &=& 0 . \label{eq:extremal_2}
\end{eqnarray}
$G^i = 0$ implies that the scalar potential reduces to the
constant term $6\, e^G$, which is consistent with the equations of
motion for the scalar fields obtained from (\ref{eq:Cremmer}) when
these fields are covariantly constant. Moreover, (\ref{eq:chi}) is
then automatically satisfied. The gravitino supersymmetry
transformation then implies that the spinorial parameter satisfies
$D_\alpha \ep + 1/2 \, e^{G/2} \gamma_\alpha \, \ep =0$.

The term $6\, e^G$ in the potential acts as a negative
cosmological term and the metric is that given in section
\ref{AdS}. (Since the scalar fields can at most be certain
functions of $u$ and will not explicitly appear in our discussion,
we shall continue to use $z$ as the real coordinate of section
\ref{AdS}). Defining the frame $e^+ = {a\over z} \, du$, $e^- =
{a\over z} \left( dv+{1\over 2} H \, du \right)$, $e^3 = {a\over
z} \, dz$, $e^4 = {a\over z} \, dx$ so that $ds^2 = 2\, e^+ e^- +
e^3 e^3 + e^4 e^4$, we find the non vanishing components of the
spin connection to be
\begin{equation}
    \begin{array}{lcl}
        \omega_{+3} &=& {1\over 2} \partial_z H \, du - {1\over a}
        e^- , \\
        \omega_{+4} &=& {1\over 2} \partial_x H \, du , \\
        \omega_{-3} &=& - {1\over z} du , \\
        \omega_{34} &=& {1\over z} dx ,
        \end{array}
\end{equation}
and the Ricci tensor can be written as
\begin{equation}
    R_{\mu\nu} = - {3\over a^2} \, g_{\mu\nu} + \delta^u_\mu \,
    \delta^u_\nu \left[ -{1\over 2} \left( \partial_x^2 H +
    \partial_z^2 H \right) + {1\over z} \partial_z H \right] .
\end{equation}
For such solutions the cosmological constant obeys $ e^G = {1\over
a^2}$ and choosing $a$ to be the positive root, we get  $a=
e^{-G/2}$. Assuming that Eqs.(\ref{eq:extremal_1}),
(\ref{eq:extremal_2}) hold, the first two supersymmetry variations
(\ref{eq:gaugino}), (\ref{eq:chi}) vanish provided $\epsilon$
satisfies $\gamma^+ \epsilon =0$ as in section
\ref{sec:Lorentzian}. Setting the variation (\ref{eq:gravitino})
of the gravitino to zero  then gives
\begin{equation}
    \begin{array}{lcl}
        \partial_v \epsilon &=& 0 , \\
        \partial_u \epsilon - {1\over 2z} \, \gamma^{-3}
        \epsilon &=& - {1\over 2z} \gamma^- \epsilon , \\
        \partial_z \epsilon &=& - {1\over 2z} \gamma^3
        \epsilon , \\
        \partial_x \epsilon + {1\over 2z} \, \gamma^{34}
        \epsilon &=& - {1\over 2z} \gamma^4 \epsilon .
    \end{array}
\end{equation}
Since $\left\{\gamma^+ \, , \, \gamma^3\right\} = 0$ we can
project these equations on the two spinors $\epsilon_1$,
$\epsilon_2$ such that $\gamma^3 \epsilon_1 = + \epsilon_1$,
$\gamma^3\epsilon_2 = - \epsilon_2$. This gives us the two systems
\begin{equation}
    \begin{array}{lr}
      \begin{array}{lcl}
        \partial_v \epsilon_1 &=& 0 , \\
        \partial_u \epsilon_1 &=& 0 , \\
        \partial_z \epsilon_1 &=& - {1\over 2z} \epsilon_1 , \\
        \partial_x \epsilon_1 &=&0 , \\
    \end{array}

   \begin{array}{lcl}
        \partial_v \epsilon_2 &=& 0 , \\
        \partial_u \epsilon_2 &=&  - {1\over z} \gamma^-
        \epsilon_2   , \\
        \partial_z \epsilon_2 &=& + {1\over 2z} \epsilon_2 , \\
        \partial_x \epsilon_2 &=& -  {1\over z} \, \gamma^{4}
        \epsilon_2  .
    \end{array}
    \end{array}
\end{equation}
The first system of equations has the solution $\epsilon_1 =
\frac{1}{\sqrt{z}} \eta$, where $\eta$ is a constant spinor
satisying $\gamma^+ \eta = 0 = ( 1 - \gamma^3 ) \eta$. The second
system of equations instead does not admit a solution. Therefore,
the presence of a cosmological constant term implies that the
solutions preserve 1/4 of the supersymmetry.

\section*{Acknowledgments} This work is supported in part by funds
provided by the U.S. Department of Energy (D.O.E.)
under cooperative research agreement DF-Fc02-94ER40818. M. C. is
supported by EPSRC and Fondazione A. Della Riccia, Florence.


\end{document}